\documentclass[fleqn,usenatbib]{mnras}

\usepackage[T1]{fontenc}
\usepackage{ae,aecompl}

\usepackage{graphicx}	
\usepackage{amsmath}	
\usepackage{amssymb}	
\usepackage{newtxtext,newtxmath}

\newcommand{\diff}{\mathrm{d}}
\newcommand{\sh}{\mathrm{sh}}
\newcommand{\view}{\theta_\mathrm{v}}
\newcommand{\thj}{\theta_\mathrm{j}}

\newcommand{\emisd}{\epsilon' _{\nu '}}
\newcommand{\numd}{\nu'_\mathrm{m}}
\newcommand{\nucd}{\nu'_\mathrm{c}}
\newcommand{\me}{m_\mathrm{e}}
\newcommand{\mpr}{m_\mathrm{p}}
\newcommand{\qe}{q_\mathrm{e}}
\newcommand{\eB}{\varepsilon_\mathrm{B}}
\newcommand{\ee}{\varepsilon_\mathrm{e}}
\newcommand{\BM}{\mathrm{BM}}

\newcommand{\Etot}{E_\mathrm{tot}}
\newcommand{\Et}{\tilde{E}_\mathrm{tot}}
\newcommand{\nt}{\tilde{n}_0}
\newcommand{\eBt}{\tilde{\varepsilon}_\mathrm{B}}
\newcommand{\eet}{\tilde{\varepsilon}_\mathrm{e}}
\newcommand{\thc}{\theta_\mathrm{c}}
\newcommand{\fb}{f_\mathrm{b}}
\newcommand{\ts}{t_\mathrm{s}}
\newcommand{\Rs}{R_\mathrm{s}}
\newcommand{\Tp}{T_\mathrm{p}}

\newcommand{\Tf}{T_\mathrm{f}}
\newcommand{\Eax}{E_\mathrm{axis}}
\defcitealias{TI}{TI20}

\title[Diverse jet structures]{Diverse Jet Structures Consistent with the Off-axis Afterglow of GRB 170817A}

\author[Takahashi \& Ioka]{
Kazuya Takahashi$^{1}$\thanks{E-mail: kazuya.takahashi@yukawa.kyoto-u.ac.jp} 
and Kunihito Ioka$^{1}$
\\
$^{1}$Center for Gravitational Physics, Yukawa Institute for Theoretical Physics, Kyoto University, Kyoto, 606-8502, Japan
}

\date{Accepted XXX. Received YYY; in original form ZZZ}

\pubyear{2020}

\begin{document}
\label{firstpage}
\pagerange{\pageref{firstpage}--\pageref{lastpage}}
\maketitle

\begin{abstract}
The jet structure of short gamma-ray bursts (GRBs) has been controversial after the detection of GRB 170817A as the electromagnetic counterparts to the gravitational wave event GW170817. Different authors use different jet structures for calculating the afterglow light curves.
We formulated a method to inversely reconstruct a jet structure from a given off-axis GRB afterglow, without assuming any functional form of the structure. 
By systematically applying our inversion method, we find that more diverse jet structures are consistent with the observed afterglow of GRB 170817A within errors: such as hollow-cone, spindle, Gaussian, and power-law jet structures. In addition, the total energy of the reconstructed jet is arbitrary, proportional to the ambient density $n_0$, with keeping the same jet shape if the parameters satisfy the degeneracy combination $n_0 \varepsilon_\mathrm{B}^{(p+1)/(p+5)} \varepsilon_\mathrm{e}^{4(p-1)/(p+5)} = \mathrm{const.}$. Observational accuracy less than $\sim 6$ per cent is necessary to distinguish the different shapes, while the degeneracy of the energy scaling would be broken by observing the spectral breaks and viewing angle.
Future events in denser environment with brighter afterglows and observable spectral breaks are ideal for our inversion method to pin down the jet structure, providing the key to the jet formation and propagation.
\end{abstract}




\section{Introduction}
The gravitational wave event GW170817 with
the detection of electromagnetic counterparts, in particular GRB~170817A and its afterglow, has revealed that
a coalescence of two neutron stars leads to a short gamma-ray burst (GRB).
The successful launch and propagation of a relativistic jet is supported by the superluminal motion of the compact radio source \citep{superluminal,Ghirlanda19} and the rapid decline of the afterglow light curve after the peak \citep{Mooley18,rapiddecline,rapiddeclineHST,Panchromatic}.
The gamma-ray emission is apparently faint because the jet is off-axis, misaligned with the line of sight
\citep[][and references therein]{170817gamma,IN18,IN19}.
In order for the jet to break out the merger ejecta,
the jet power should be similar to the other short GRBs
\citep{Nagakura14,Hamidani19}.

GRB~170817A also revealed that the jet has to be structured with an angular dependent energy distribution, which is required to explain the slowly rising light curves of the afterglow \citep{Mooley18}. However, the exact jet structure is still controversial. Different authors assume different structures for calculating the afterglow light curves. Often used structure is a Gaussian \citep{ZM02,Rossi04,Lyman18,Resmi18,Troja19,LK18,rapiddeclineHST,GG20,Ryan20,Troja20} or a power law \citep{Meszaros98,Rossi02,Rossi04,ZM02,GK03,LK18,DAvanzo18,Ghirlanda19,Beniamini20,GG20,Ryan20}, although there is no strong motivation to limit the jet structure to these specific forms. 
The jet structure is crucial to off-axis gamma-ray emission \citep{Beniamini19,IN19}. Furthermore, the jet structure is potentially a clue to the jet launching, collimation, and propagation processes as inferred from numerical simulations \citep{Aloy05,Nagakura14,Duffell15b,Lazzati17,MB17,Kathirgamaraju18,Kathirgamaraju19,Xie18,Geng19,Gill19,Gottlieb20,Nathanail20}.

We recently discovered that a hollow-cone jet structure is also a candidate for GRB~170817A for the first time \citep[][hereafter \citetalias{TI}]{TI}. It was discovered by a new approach that inversely reconstructs a jet structure
from a given afterglow light curve in off-axis GRBs. One of the advantages of the method is that it does not prefix a functional form of the jet structure. Thus, as demonstrated by the discovery of a hollow-cone jet, the method can reconstruct a non-trivial jet structure that is different from a Gaussian jet or a power-law one. Another advantage is that the method uniquely determines a jet structure for a given light curve, once the other model parameters are fixed.\footnote{We use the term ``unique'' related to the jet structure under the assumption of a fixed edge structure throughout the paper. As stated later, our method should assume the jet edge structure, which cannot be determined from the rising part of afterglows. The jet inner structure quantitatively depends on the assumed edge structure, but qualitatively remains the same as shown in \citetalias{TI}.} 
The hollow-cone jet is reconstructed from a straight-line light curve, which is the simplest example that fits the rising slope of the observed afterglow. In addition, the hollow-cone type structure is always reconstructed by using the straight-line light curve, irrespective of the other model parameters such as the viewing angle and the jet edge structure while the jet structure changes quantitatively. This result indicates that the light curve shape is the paramount feature for determining the jet structure. However, in reality, we could draw another light curve to fit the observed data within the uncertainty, which leads to a Gaussian jet or a power-law one depending on the light curve shape. So far, 
it has not been studied thoroughly how the jet structure changes for different light curves.

This paper searches for jet structures that are consistent with the afterglow of GRB~170817A in a systematic way.
We apply the inversion method in \citetalias{TI} to more general types of the light curves by taking into account the curvature of the light curve. The light curve shape is changed in a parametric way so that it is consistent with the afterglow of GRB~170817A. We find that a little difference of the light curve creates a large diversity of non-trivial jet structures.
We also find that the total energy of the jet is not determined because we can change the total energy without changing the jet shape by appropriately tuning the afterglow parameters.

The paper is organized as follows. In Section~\ref{sec.method}, we introduce the parametric light curve and other model parameters used in this study after briefly summarizing the inversion method in \citetalias{TI}. Then, the various jet structures reconstructed from the generic light curves are presented in Section~\ref{sec.results}. We discuss the reason of the diversity of the jet candidates in Section~\ref{sec.discussion}. Finally, we make a conclusion on the jet structure of GRB~170817A in Section~\ref{sec.conclusion}. Throughout the paper, we attach a prime to the quantities evaluated in the fluid rest frame.

\section{Method}\label{sec.method}
\subsection{Brief Review of the Inversion Method}\label{sec.dof}
We consider an axi-symmetric relativistic jet propagating into the interstellar medium with a constant number density $n_0$. The jet energy has an angular structure $E(\theta)$, where $E$ is the isotropic equivalent energy and $\theta$ is the polar angle measured from the jet axis.
A shock is formed in the interstellar medium, and is well described by a self-similar solution of \citet{BM} 
as if it were a portion of a spherical blast wave until the shock is further decelerated to non-relativistic speeds \citep{KG03,ZM09,EM12}.
The afterglow is produced by synchrotron radiation emitted from 
electrons accelerated to a power-law energy distribution with a spectral index $p$ 
at the shock with a magnetic field in the downstream \citep{Sari98}.
The shocked region is approximated by a thin shell \citep{Eerten10}.
Then, we can calculate the afterglow light curve for a given jet energy structure $E(\theta)$. 

Inversely, we can reconstruct the jet structure $E(\theta)$ from a given afterglow light curve in the case of off-axis GRBs. 
This is possible 
because the jet structure
gradually becomes visible from the jet edge to the central region,
as the relativistic beaming angle increases due to the deceleration of the jet
sweeping the ambient material. The emission centroid gradually moves from the jet edge region close to the observer direction toward the jet axis, scanning the jet structure as time passes and reflecting the jet structure into the light curve.

Based on the above idea, we recently formulated an inversion method that reconstructs a jet structure from an afterglow light curve \citepalias{TI}. While the observed light curve is calculated by integrating the synchrotron emission radiated from the shock surface, 
we reverse the integral into an ordinary differential equation 
for $E(\theta)$ in Equation~(\ref{eq.inversion}) below. By integrating Equation~(\ref{eq.inversion}), the jet structure is uniquely determined from a given afterglow light curve. Since the scan of the jet surface is finished when the emission site reaches the jet axis, the reconstruction method uses only a part of the rising portion of the light curve before the peak.

The inversion formula is given by (See Appendix~\ref{sec.review} for a short review of the derivation and the detail)
\begin{align}
\label{eq.inversion}
\frac{\diff \ln E}{\diff \Theta} &= \frac{8}{\view - \Theta} -\frac{3 K(T,\Theta,E(\Theta))}{F_\nu(T)} \left[\frac{\diff \log F_{\nu}}{\diff \log T}(T) \right. \nonumber \\
&\qquad \qquad \qquad \quad \left. - \frac{T}{F_\nu(T)}\int _{\Theta}^{\thj} \diff \theta \frac{\diff K}{\diff T}(T,\theta,E(\theta))\right]^{-1},
\end{align}
where $F_\nu(T)$ is the observed flux with $\nu$ and $T$ being the observed frequency and observer time, respectively. 
$\view$ is the viewing angle measured from the jet axis.
$\Theta(T)$ approximately corresponds to the polar angle of the emission centroid and satisfies Equation~(\ref{eq.Theta}).
$K$ is a function defined by Equation~(\ref{eq.K}), and $\thj$ is the jet truncation angle.
A jet structure is obtained by integrating Equation~(\ref{eq.inversion}) from $\Theta=\Theta(T_0)$ to $\Theta=0$ for a given initial observer time $T_0$.
The effect of the cosmological redshift is neglected for simplicity.
Necessary ingredients are the following:
\begin{enumerate}
	\item The rising portion of the light curve $F_\nu(T)$ ($T_0 \le T \le \Tf$) for a given observed frequency $\nu$. Here, $T_0$ is a given initial time. $\Tf$ is the time when $\Theta(T)$ reaches the jet axis and the reconstruction is completed, 
	which is given by Equation~(\ref{eq.Tf}). Since $\Tf$ depends on the isotropic equivalent energy at the jet axis, $\Eax$, it is not \textit{a priori} known until the end of the inverse reconstruction.
	
	\item Model parameters $\{n_0, \eB, \ee, \view, p, D\}$, where $n_0$ is the number density of the ambient interstellar medium, $\eB$ the energy conversion fraction from the shocked matter to the magnetic field, $\ee$ the energy conversion fraction from the shocked mater to the accelerated electrons, $\view$ the viewing angle, $p$ the energy spectral index of the accelerated electrons, and $D$ the luminosity distance to the source.
	
	\item Jet edge structure $E(\theta)$ for $\Theta(T_0) \le \theta \le \thj$ with $\thj$ being a jet truncation angle. This is given as a boundary condition for integrating Equation~(\ref{eq.inversion}). The jet edge structure cannot be 
    constrained from the light curve after $T_0$
    in principle, since this region has been already scanned before $T_0$.
\end{enumerate}

This paper explores the allowed range of the jet structure by focusing on the uncertainty of the light curve listed as (i) above.
The observed data points generally include error bars and distribute sparsely in time. In the case of GRB~170817A, the rising slope could be fitted by a single power law $F_\nu \propto T^\alpha$ 
with $\alpha$ being a constant, while the slope may have a curvature.
\citetalias{TI} used a simple power-law light curve, $\diff \log F_\nu /\diff \log T = 1.22 = \mathrm{const.}$, and found that it always reconstructs qualitatively the same hollow-cone jet structure irrespective of the other parameters. Hence, it is a natural extension to 
consider more general light curves. The light curve used in this paper is given in Section~\ref{sec.LCdep}.

In the rest of this subsection, we comment on the uncertainty of the other factors listed in (ii) and (iii) above. As mentioned later in Section~\ref{sec.degeneracy_main}, some of the model parameters listed in (ii) are degenerate and only a few are constrained from observations in general. 
In the case of GRB~170817A, the spectral index $p\sim 2.17$ \citep{Fong19,Troja19} and luminosity distance $D \sim 41$~Mpc \citep{D,Cantiello18} are relatively well constrained while the others are not:
The viewing angle $\view$ is constrained from the detection of a super-luminal motion of a compact radio source in the afterglow phase \citep[$14^\circ \lesssim \view \lesssim 28^\circ$;][]{superluminal}. 
Note that the uncertainty of the viewing angle leads to the uncertainty of the jet core width, $\theta_\mathrm{c}$. We reconstruct a wider jet structure if we assume a larger $\view$ \citepalias{TI}. 
It is consistent with \citet{NP20}, who pointed out that the degeneracy between $\view$ and $\thc$ is not broken from the afterglow light curve alone.
\footnote{\citet{NP20} showed that $\view/\thc$ is insensitive to the rising part of light curves and inferred from the width of the peak. Our inversion method determines the jet structure (i.e.. $\thc$) from the rising part for a given $\view$ {\it only if} the rising part is given very accurately. As shown in Section~\ref{sec.results}, even a small difference in the rising part leads to a different jet structure (i.e., a different ratio of $\view/\thc$). Hence, our finding is consistent with \citet{NP20}.}
$n_0$ is constrained in several ways: The diffuse X-ray emission from the host galaxy puts a constraint of $n_0 \le 9.6\times 10^{-3}$~cm${}^{-3}$ \citep{Hajela19} while the non-detection of neutral hydrogen from the host also gives a looser upper limit $n_0 \le 4 \times 10^{-2}$~cm${}^{-3}$ \citep{Hallinan17}.
On the other hand, a cosmological simulation infers a lower limit of $n_0 \ge 2 \times 10^{-5}$~cm${}^{-3}$ \citep{Shull,Mooley18}.
$\eB$ and $\ee$ have not been constrained from the observations so far, whereas $\ee=0.1$ is a typical value in observations of other GRBs \citep{KZ15} and in particle-in-cell simulations \citep{SS11}. 

As noted in (iii), the jet edge structure is not constrained in principle from afterglow observations. However, the jet edge structure would be less important for the reconstruction of the inner part close to the jet axis, since the light curve for later time is dominated by the emission from the inner part. 
As shown in \citetalias{TI}, 
the inner jet structure is qualitatively determined irrespective of whether the assumed edge structure is a Gaussian or a power-law.
Especially, the jet truncation angle $\thj$ does not change the results qualitatively nor quantitatively as long as we choose a sufficiently large value \citepalias{TI}.

\subsection{Parametric Light Curve and Fiducial Setting}\label{sec.LCdep}
In the previous paper \citepalias{TI}, 
we simplified the rising phase of the afterglow of GRB~170817A as a light curve with a constant slope (that is, the logarithmic derivative $\diff \log F_\nu / \diff \log T$ is constant).
In this paper, we use more general light curves for the rising phase of the afterglow. 

As a generalization,
we consider the light curves that are parametrically given by
\begin{align}
	\label{eq.LC_generalized}
	F_\nu(T) = F_0\left(\frac{T}{T_0}\right)^{-aT_0 + b} \exp[a(T-T_0)] \ (T_0 \le T \le \Tf),
\end{align}
where $a$ is the curvature parameter, which controls the changing rate of the light curve slope, and $b$ is the initial slope as seen below.
$T_0$ is the initial observer time and $F_0=F_\nu(T_0)$ is the corresponding initial flux. 
The slope of the light curve in Equation~(\ref{eq.LC_generalized}) is
\begin{align}
\label{eq.LCslope_generalized}
\frac{\diff \log F_\nu}{\diff \log T} = a(T - T_0) + b.
\end{align}
Thus, the light curve slope is initially given by $\diff \log F_\nu / \diff \log T = b$ at $T=T_0$ while it changes as time passes with a rate of $a$: In ($\log F_\nu$, $\log T$) plane, $a>0$ gives convex downward light curves, $a<0$ gives convex upward ones, and $a=0$ gives straight ones. We note that the case with $a=0$ and $b=1.22$ was investigated in \citetalias{TI}, where a hollow-cone jet is inversely reconstructed.

Changing the curvature parameter $a$, we reconstruct the jet structure
while we fix $\nu=5.5$~GHz, $b=1.22$, $T_0=9$~d, and $F_0=5.45$~$\mu$Jy in Equation~(\ref{eq.LC_generalized}).
We use the fiducial model parameters as $\ee=0.1$, $p=2.17$, $\view=0.387 \sim 22.2^\circ$, and $D=41$~Mpc, while we treat $n_0$ and $\eB$ as free parameters to adjust the peak time and peak flux of the light curve that is forwardly calculated by the reconstructed jet. Note here that, since the jet structure is reconstructed from the rising portion of the light curve with $T_0 \le T \le \Tf$ before the peak, we check as a post process whether the light curve forwardly calculated from the reconstructed jet agrees with the observations for $T > \Tf$. We repeat the reconstruction procedure by iteratively tuning $n_0$ and $\eB$ until the forwardly calculated light curve matches the observed one. For the sake of 
quantitative accuracy,
we use a more accurate afterglow equation, equation~(18) in \citetalias{TI}, to forwardly calculate the light curves than Equation~(\ref{eq.Fapp}), which is equivalent to the inversion formula, Equation~(\ref{eq.inversion}). 
The accurate afterglow equation takes into account the flux contribution from the entire jet surface, a smooth transition of the shock dynamics from the Blandford-McKee solution to the Sedov-Taylor solution \citep{Taylor,Sedov,Eerten10}, the spectral breaks at the characteristic and cooling frequencies \citep[$\numd$ and $\nucd$, respectively;][]{Sari98}, and so on. The relative difference between the light curves calculated by the accurate equation and Equation~(\ref{eq.Fapp}) is within $5$ per cent.\footnote{The discrepancy between the light curves calculated by the accurate
equation and Equation~(\ref{eq.Fapp}) in this paper is mainly attributed to the inner cutoff angle $\Theta(T)$ in Equation~(\ref{eq.Fapp}), which ignores the emission from the region near to the jet axis, $0 \le \theta < \Theta(T)$. This discrepancy is inevitable, since $\Theta(T)$ plays an essential role in the inversion method. However, we can reduce it, if necessary, by using larger $\fb$ in Equation~(\ref{eq.Theta}), which gives smaller $\Theta(T)$ for a fixed $T$ (See \citetalias{TI} for details). \label{ftnt.used_and_forward}}

We assume the jet edge structure described by a Gaussian shape, which is given by
\begin{align}
\label{eq.GaussianJet}
E(\theta) = E_0 \exp\left(-\frac{\theta^2}{2\thc^2}\right)\ (\Theta (T_0) \le \theta \le \thj),
\end{align}
where we use $\thj=0.61 \sim 35.0^\circ$ as our fiducial value although the choice of $\thj$ does not affect the results as long as it is large enough as mentioned in Section~\ref{sec.dof}. We also note that the inner jet structure does not change qualitatively with the choice of the edge shape as noted before.
$E_0$, $\thc$, and $\Theta(T_0)$ are self-consistently determined so that the edge structure becomes consistent with given $F_\nu(T_0)$ and $\diff F_\nu/\diff T(T_0)$ for each light curve (See Appendix~\ref{app.jetedge}).

\subsection{Degeneracy Relation in $(n_0, \eB, \ee) $}\label{sec.degeneracy_main}
In the inversion process for a given light curve, 
we generally find that the jet structures with the same shape are reconstructed from different combinations of $(n_0, \eB, \ee)$.
Here, we mean by the same shape that the jet structures have the same angle dependence, $f(\theta)$, except for the normalization, $\Eax$, as
\begin{align}
    \label{eq.shape}
    E(\theta) = \Eax f(\theta)
\end{align}
with $f(0)=1$ and $\Eax = E(0)$.

As shown in Appendix~\ref{sec.degen}, the degenerate combinations of $(n_0, \eB, \ee)$ are given by the following equation
\begin{align}
\label{eq.degen1}
\log n_0 + \frac{p+1}{p+5} \log \eB + \frac{4(p-1)}{p+5}\log \ee = \mathrm{const.} =: \log \xi,
\end{align}
where $\xi$ is the degeneracy parameter that characterizes the combination. 
For the degenerate combinations of $(n_0, \eB, \ee)$, which satisfy Equation~(\ref{eq.degen1}) for a given $\xi$, the reconstructed jet structures have the same shape $f(\theta)$ while the energy $\Eax$ is scaled by $n_0$:
\begin{align}
\label{eq.degen2}
\Eax \propto n_0.
\end{align}
In other words, the parameters satisfying Equations~(\ref{eq.degen1}) and (\ref{eq.degen2}) give the same light curve. The
total energy of the (one-sided) jet $\Etot$ is also scaled as $\Etot \propto \Eax \propto n_0$, where $\Etot$ is given by
\begin{align}
\label{eq.Etot}
\Etot = 2\pi \int _0 ^{\thj} E(\theta) \sin \theta \diff \theta = 2\pi \Eax \int _0 ^{\thj} f(\theta) \sin \theta \diff \theta \propto n_0.
\end{align}

By using Equations~(\ref{eq.degen1}) and (\ref{eq.Etot}), we can rescale the total energy of a jet to a desired value without changing the afterglow light curve. For example, suppose that we have a jet structure with a total energy $\Etot$ that is reconstructed from a light curve with a parameter set $(n_0, \eB, \ee)$. Then, if we change the total energy to a desired value $\Et$, the corresponding number density $\nt$ is given by Equation~(\ref{eq.Etot}) as follows:
\begin{align}
    \label{eq.nt}
    \nt = \frac{\Et}{\Etot}n_0.
\end{align}
According to Equations~(\ref{eq.degen1}) and (\ref{eq.nt}), the appropriate combinations of the microphysical parameters, $(\eBt,\eet)$, should satisfy
\begin{align}
    \label{eq.et}
    \eBt \eet^{4(p-1)/(p+1)} = \eB \ee^{4(p-1)/(p+1)} \left(\frac{\Etot}{\Et}\right)^{(p+5)/(p+1)}.
\end{align}

Not only the peak time and flux but also the whole part of the rising light curve remain the same when the afterglow parameters satisfy the degeneracy relations, Equations~(\ref{eq.degen1}) and (\ref{eq.degen2}). We also emphasize that the degeneracy relations are applicable to off-axis GRBs with any structured jet. This is a generalization of the previous findings: \citet{NP18} showed the same dependence of the light curve on the parameters ($n_0$, $\eB$, $\ee$) for the on-axis GRBs [their equation (1)] and \citet{Gill19} also reached essentially the same degeneracy relations from the invariance of the peak time and flux [their equation (7)]. The degeneracy between some parameters is mentioned in the case of GRB~170817A \citep{superluminal,Beniamini20}. We also comment that the parameter estimation using the spectral breaks in afterglows has been well known \citep{WG99}, and no detection of breaks in the GRB~170817A afterglow is a cause of the parameter degeneracy.

\begin{figure*}
	\includegraphics[width = \textwidth]{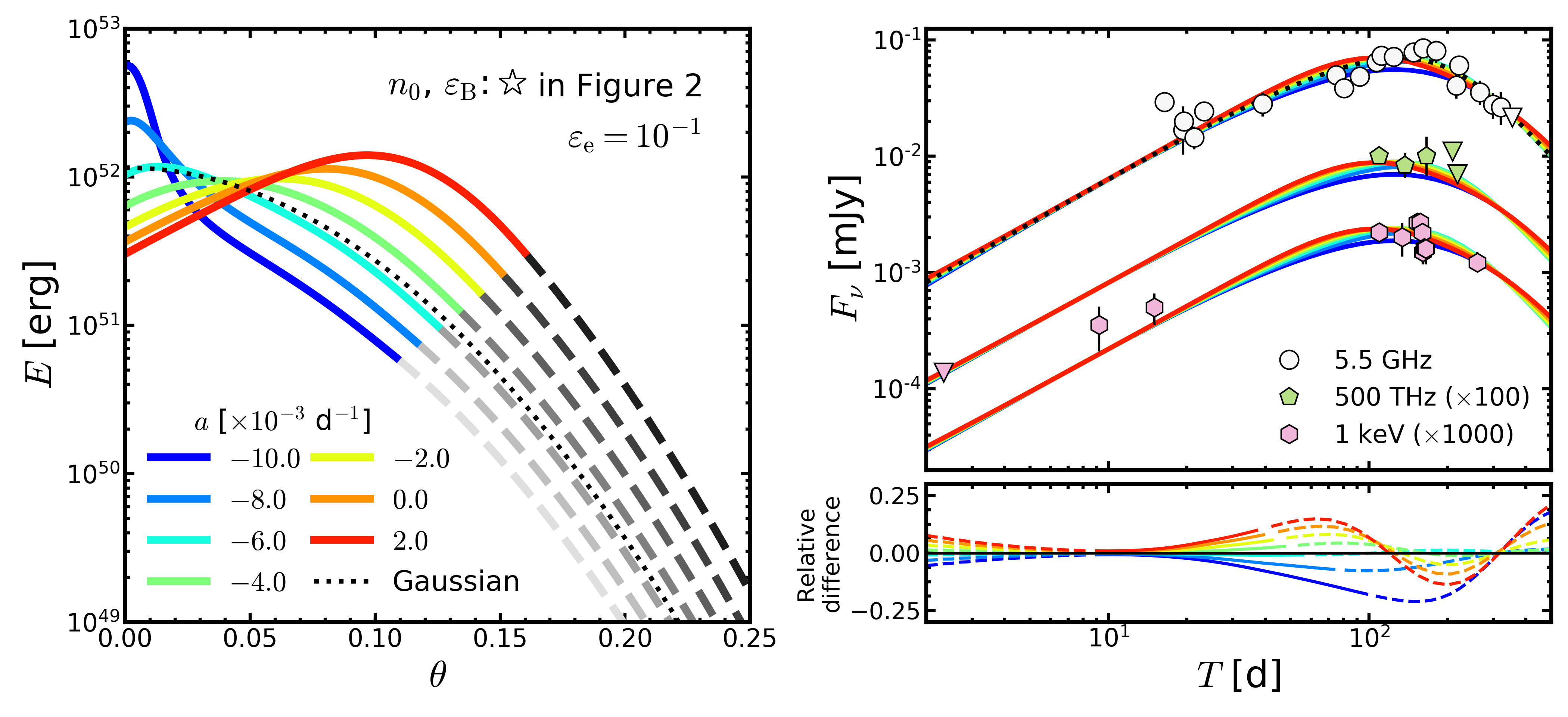}
	\caption{
	{\bf Left:} Jet structures inversely reconstructed from the radio ($\nu=5.5\ \mathrm{GHz}$) light curves with different values of the curvature parameter $a$. The light curves used for inversion are given by Equation~(\ref{eq.LC_generalized}) with $-10\le a/(10^{-3}\ \mathrm{d}^{-1}) \le 2$, $b=1.22$, $T_0=9\ \mathrm{d}$, and $F_0 = 5.45\ \mu\mathrm{Jy}$. The values of $n_0$ and $\eB$ are listed in Table~\ref{tab.result_star} and are also shown by a star-shaped point in Figure~\ref{fig.eBn0} for each model, while the other parameter values are fiducial ones: $\ee=10^{-1}$, $\view = 0.387$, $p=2.17$, and $D=41$~Mpc. The coloured solid lines show the inversely reconstructed jet structures ($\theta < \Theta(T_0)$) while the gray dashed lines correspond to the jet edge part given by Equation~(\ref{eq.GaussianJet}) with $\thj=0.61$ ($\theta \ge \Theta(T_0)$), which has to be assumed for inversion. 
	The black dotted line shows the Gaussian given by $E = E_\mathrm{G} \exp[-\theta^2/(2\theta_\mathrm{G}^2)]$ with $E_\mathrm{G}=1.16 \times 10^{52}\ \mathrm{erg}$ and $\theta_\mathrm{G}=5.90\times 10^{-2}$.
	{\bf Top Right:} Light curves forwardly calculated by the reconstructed jets shown in the left panel, which are for radio (5.5~GHz), optical (500~THz), and X-ray (1~keV) from top to bottom for each colour.
	The black dotted line shows the radio light curve that is generated by the Gaussian jet with $(n_0, \eB)$ presented by the black star in Figure~\ref{fig.eBn0}.
    Also shown for reference are the observed afterglow (points) and upper limits (lower triangles) of GRB~170817A. 
	The data points for radio were taken from Figure~4 in \citet{Troja19}, which uses the data in \citet{Hallinan17,Lyman18,Troja18a,Margutti18,Mooley18,Alexander18,Piro19}. The data points for optical and X-ray were collected from \citet{Lyman18,Margutti18,DAvanzo18,Alexander18,Piro19}.
	{\bf Bottom Right:} Relative difference of the forwardly calculated light curve with respect to the Gaussian jet light curve, which is defined by $[F_\nu(\mathrm{forward}) - F_\nu(\mathrm{Gaussian})]/F_\nu(\mathrm{Gaussian})$. The solid lines are used for $T_0 \le T \le \Tf$ while the dashed lines are used for others.
	}
	\label{fig.result_star}
\end{figure*}

\begin{table}
	\centering
	\caption{Parameters and results for Figure~\ref{fig.result_star}, where 
		$a_{-3} = a/(10^{-3}\ \mathrm{d}^{-1})$ and
		$Q_x = Q/10^x$ in cgs units for the other quantities.
		$a$: the curvature parameter of the light curve in Equation~(\ref{eq.LC_generalized}). 
		$n_0$: the number density of the ambient matter. 
		$\eB$: the energy conversion fraction from the shocked matter to the magnetic field. 
		$\xi$: the degeneracy parameter in Equation~(\ref{eq.degen1}). 
		$E_0$ and $\thc$: the normalization and standard deviation of the Gaussian edge structure in Equation~(\ref{eq.GaussianJet}), respectively. 
		$\Etot$: the total energy of the (one-sided) jet given by Equation~(\ref{eq.Etot}).
	}
	\label{tab.result_star}
	\begin{tabular}{ccccccc}
		\hline
		$a_{-3}$ & $n_{0,-3}$ & $\varepsilon_{\mathrm{B},-4}$ & $\xi_{-6}$ & $E_{0,52}$ & $\theta_{\mathrm{c},-2}$ & $E_{\mathrm{tot},50}$ \\
		\hline
		$-10$ &  1.00 & 6.60 & 8.73  & 0.34 & 5.83  & 1.12  \\
		$-8$  &  1.00 & 4.41 & 7.31  & 0.56 & 5.86  & 1.60  \\
		$-6$  &  1.00 & 2.87 & 6.04  & 0.94 & 5.89  & 2.26  \\
		$-4$  &  1.00 & 1.85 & 4.98  & 1.64 & 5.91  & 3.08  \\
		$-2$  &  1.00 & 1.17 & 4.06  & 2.96 & 5.92  & 4.18  \\
		$0$   &  1.00 & 0.71 & 3.26  & 5.70 & 5.92  & 5.80  \\
		$2$   &  1.00 & 0.41 & 2.56  & 11.8 & 5.92  & 8.19  \\
		\hline
	\end{tabular}
\end{table}

\begin{figure}
	\includegraphics[width = \columnwidth]{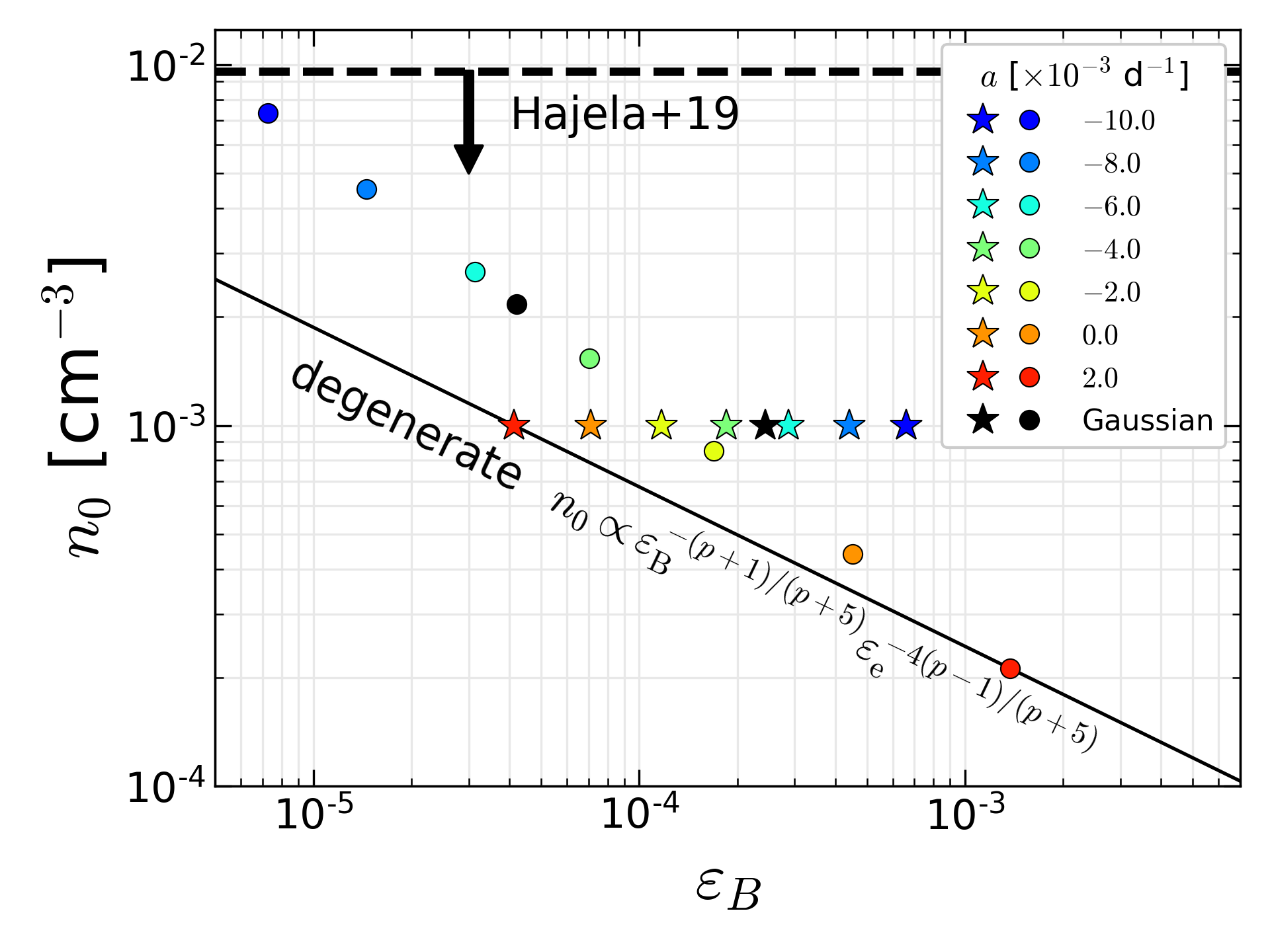}
	\caption{The number density of the ambient matter $n_0$ and the energy conversion fraction from the shocked matter to the magnetic field $\eB$ used for the results in Figure~\ref{fig.result_star} (star-shaped points) and those for Figure~\ref{fig.result_bullet} (round-shaped points). Also shown is the upper limit of $n_0$ given by \citet{Hajela19}. Note that the pairs of $(n_0, \eB)$ for the same-coloured star and round points lie on a line that satisfies the degeneracy relation $n_0 \propto \eB^{-(p+1)/(p+5)} \ee^{-4(p-1)/(p+5)}$ [Equation~(\ref{eq.degen1})], as indicated by the black solid line for the red points for example. The pairs $(n_0, \eB)$ in the degeneracy relation give the jet structures with the same shape but with the total energy that scales with $n_0$ for a given afterglow light curve.}
	\label{fig.eBn0}
\end{figure}

\begin{figure*}
    \includegraphics[width = \textwidth]{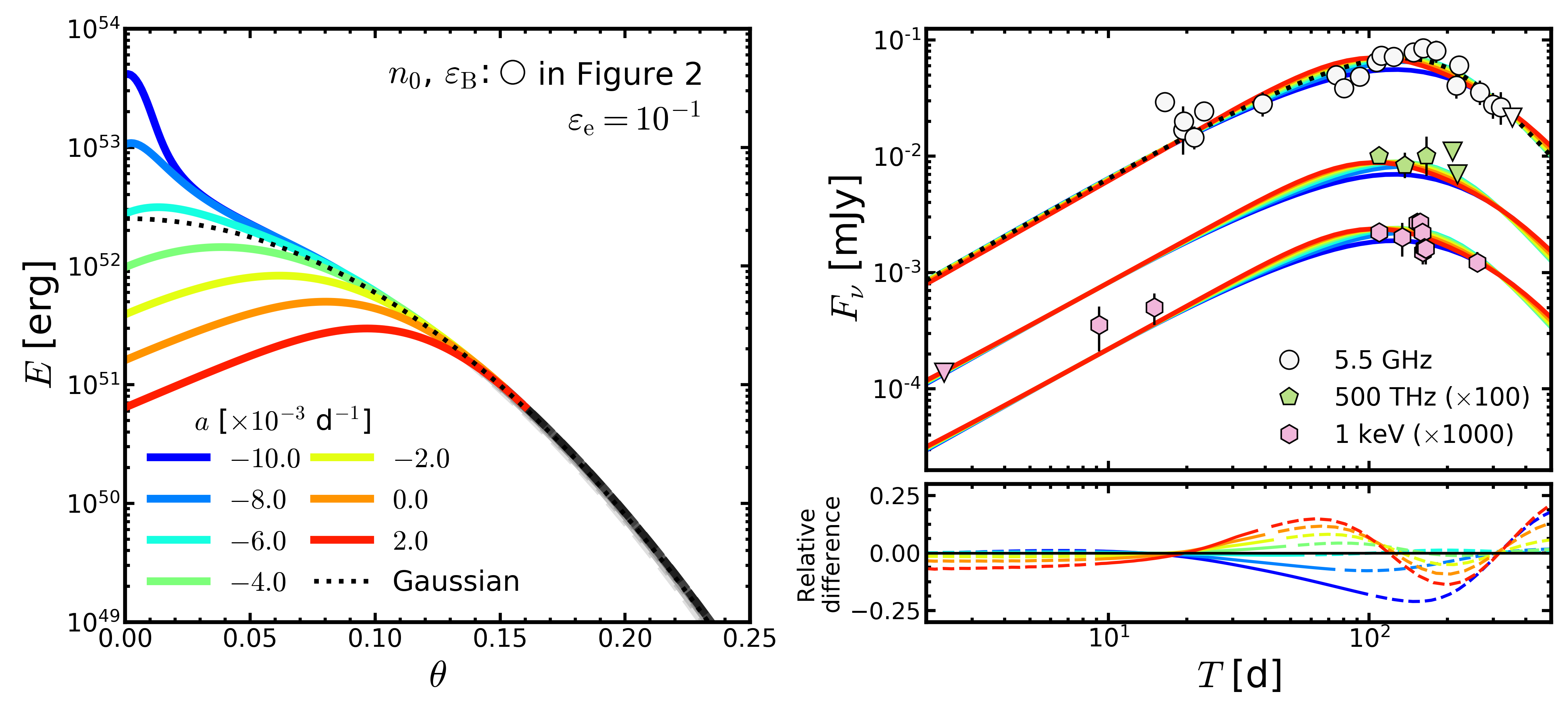}
	\caption{Same as Figure~\ref{fig.result_star} but for $n_0$ and $\eB$ that are listed in Table~\ref{tab.result_bullet} and are also shown by the round-shaped points in Figure~\ref{fig.eBn0}. The dotted line in the left panel shows the Gaussian given by $E = E_\mathrm{G} \exp[-\theta^2/(2\theta_\mathrm{G}^2)]$ with $E_\mathrm{G}=2.51 \times 10^{52}\ \mathrm{erg}$ and $\theta_\mathrm{G}=5.90\times 10^{-2}$.
	The jet structure reconstructed from each light curve has the same shape as the counterpart in Figure~\ref{fig.result_star} but has a different total energy. 
	The forwardly calculated light curves in optical and X-ray agree well with those in Figure~\ref{fig.result_star} while the radio light curve is slightly (at most $10$ per cent) different (See also Figure~\ref{fig.app_reladiff}).
	}
	\label{fig.result_bullet}
\end{figure*}

\begin{table}
	\centering
	\caption{
	    Same as Table~\ref{tab.result_star} but for Figure~\ref{fig.result_bullet}.
	    The pair of $(n_0, \eB)$ for each model is chosen so that it gives $E_{0,52}=2.51$ and the same $\xi$ as in Table~\ref{tab.result_star}. $E_{\mathrm{tot},50}/n_{0,-3}$ in the rightmost column is to coincide
        with $E_{\mathrm{tot},50}$ in Table~\ref{tab.result_star}.
	}
	\label{tab.result_bullet}
	\begin{tabular}{cccccccc}
		\hline
		$a_{-3}$ & $n_{0,-3}$ & $\varepsilon_{\mathrm{B},-4}$ & $\xi_{-6}$ & $E_{0,52}$ & $\theta_{\mathrm{c},-2}$ & $E_{\mathrm{tot},50}$ & $\displaystyle \frac{E_{\mathrm{tot},50}}{n_{0,-3}}$\\
		\hline
		$-10$ &  7.35 & 0.073 & 8.73 & 2.51 & 5.83 & 8.25 & 1.12 \\
		$-8$  &  4.52 & 0.15  & 7.31 & 2.51 & 5.86 & 7.26 & 1.60 \\
		$-6$  &  2.66 & 0.31  & 6.04 & 2.51 & 5.89 & 6.02 & 2.26 \\
		$-4$  &  1.53 & 0.70  & 4.98 & 2.51 & 5.91 & 4.72 & 3.08\\
		$-2$  &  0.85 & 1.69  & 4.06 & 2.51 & 5.92 & 3.55 & 4.18 \\
		$0$   &  0.44 & 4.53  & 3.26 & 2.51 & 5.92 & 2.55 & 5.80 \\
		$2$   &  0.21 & 13.8  & 2.56 & 2.51 & 5.92 & 1.74 & 8.19 \\
		\hline 
	\end{tabular}
\end{table}

\begin{figure}
	\includegraphics[width = \columnwidth]{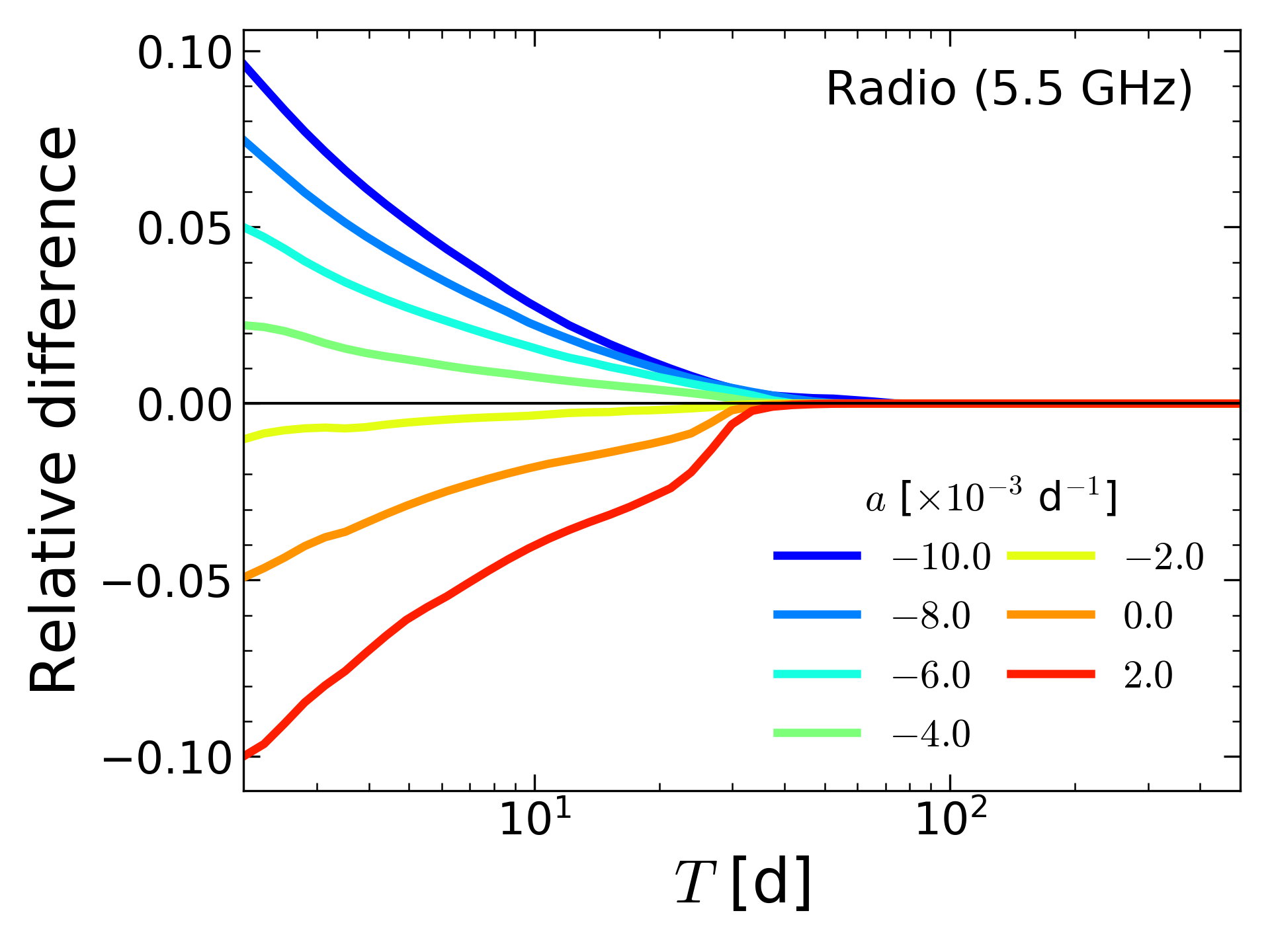}
	\caption{
		Relative difference of the forwardly calculated radio light curves in Figures~\ref{fig.result_star} and \ref{fig.result_bullet} for each $a$, which is defined by $[F_\nu(\mathrm{Fig.~3}) - F_\nu(\mathrm{Fig.~1})]/F_\nu(\mathrm{Fig.~1})$.
		The light curves are slightly different because of the different contribution from the region close to the jet axis, where the frequency drops below the synchrotron characteristic frequency in the fluid rest frame, $\nu' < \numd$, for which the emissivity increases with $n_0$. As the shock is decelerated, $\nu'  > \numd$ is realized in the entire region of the shock and the difference is reduced.
	}
	\label{fig.app_reladiff}
\end{figure}

\section{Results}\label{sec.results}
We reconstruct the jet structures 
by systematically changing the light curve slope. We use the light curve given by Equation~(\ref{eq.LC_generalized}) and control the slope by changing the curvature parameter $a$. We determine the ambient number density $n_0$ and microphysical parameter $\eB$ by adjusting the peak of the light curve that is forwardly calculated by the reconstructed jet while the other parameters are fixed to the fiducial values mentioned in Section~\ref{sec.LCdep}.

In the first part, we discover various non-trivial jet structures that are consistent with GRB~170817A within the observational errors.
The totally different structures are found to produce very similar light curves.
In the second part, we demonstrate that the total energy can be altered
without changing the structural shape nor the light curve by appropriately choosing a pair of $(n_0, \eB)$.

\subsection{Various Non-trivial Jet Structures}\label{sec.results_n0-3}
We here use the light curves with the curvature parameter $a$ in the range of $-10\le a/(10^{-3}\ \mathrm{d}^{-1}) \le 2$. We fix $n_0 = 10^{-3}$~cm${}^{-3}$ and tune $\eB$ for each light curve to adjust the peak. Figure~\ref{fig.result_star} shows the results, with $\eB$ in Table~\ref{tab.result_star} and also in Figure~\ref{fig.eBn0} (star-shaped points).

The left panel of Figure~\ref{fig.result_star} shows the reconstructed jet structures. As $a$ increases, the reconstructed jet structure continuously changes from a spindle type ($a \lesssim -6 \times 10^{-3}\ \mathrm{d}^{-1}$) to a hollow-cone type ($a \gtrsim -6 \times 10^{-3}\ \mathrm{d}^{-1}$) by passing through a Gaussian-like structure ($a \sim -6 \times 10^{-3}\ \mathrm{d}^{-1}$). For reference, we also depict a Gaussian jet in the left panel of Figure~\ref{fig.result_star} (black dotted line). The standard deviations of the Gaussian edge are nearly the same among the models, $\thc \sim 5.9 \times 10^{-2}$, while the normalization, $E_0$, is different among the models as listed in Table~\ref{tab.result_star}. The total energy $\Etot$, which is defined by Equation~(\ref{eq.Etot}), increases for larger $a$ as in Table~\ref{tab.result_star}.

In spite of the difference of the jet structures, they generate almost similar light curves, which are consistent with the observations as shown in the top right panel of Figure~\ref{fig.result_star}.
We compare these light curves quantitatively with the Gaussian jet case (dotted lines). 
The black dotted line in the top right panel shows the light curve for the Gaussian jet, which is produced with $\eB=2.44 \times 10^{-4}$ while the other parameter values are the same as for the other models. The bottom right panel shows the relative difference with respect to the light curve for the Gaussian jet. 
The relative difference increases with $T$ during the time interval that is
used for the inversion, $T_0 \le T \le \Tf$ (solid lines). 
It reaches $\sim 18$ per cent at $\Tf = 97.9$~d in the case of the spindle jet for $a=-10\times 10^{-3}$~d$^{-1}$. This means that at least the observational accuracy less than $\sim 18$ per cent is necessary for distinguishing the spindle jet from the Gaussian jet. 
The light curve for the later phase ($T > \Tf$) cannot be used for inversely reconstructing the jet structure, since the jet axis has been already seen to the observer. 
In the case of the hollow-cone jet for $a=2\times 10^{-3}$~d$^{-1}$, the relative difference of the light curves reaches $\sim 5.8$ per cent at $\Tf = 37.0$~d. Hence, the observational precision less than $5.8$ per cent is necessary to distinguish the hollow-cone jet from the Gaussian jet from the viewpoint of the inverse reconstruction.
In the multiband observations of GRB~170817A, the one-sigma errors are comparable or larger than the above demanded values in the rising phase \citep{Panchromatic}.

\subsection{Jet Structure with the Same Shape but with a Different Total Energy}\label{sec.results_another}
We demonstrate here that the same light curve could be explained by different jet structures that have the same shape but have different total energy, if the ambient number density $n_0$ and microphysical parameter $\eB$ are tuned appropriately for each case.
We compare the results with those in Section~\ref{sec.results_n0-3}.
It is shown that the total energy of the reconstructed jet is proportional to $n_0$, as expected from Section~\ref{sec.degeneracy_main}.

Figure~\ref{fig.result_bullet} shows the results obtained for the light curves given by Equation~(\ref{eq.LC_generalized}) with the curvature parameter $a$ in the range of $-10\le a/(10^{-3}\ \mathrm{d}^{-1}) \le 2$, which are the same as in Section~\ref{sec.results_n0-3}, and newly tuned $n_0$ and $\eB$, which are shown in Table~\ref{tab.result_bullet} and in Figure~\ref{fig.eBn0} (round-shaped points). The parameters are tuned by using the degeneracy relation in Section~\ref{sec.degeneracy_main}: $n_0$ is given by Equation~(\ref{eq.nt}) so as to have $E_0=2.51\times 10^{52}$ for the normalization of the jet edge and $\eB$ is accordingly given by Equation~(\ref{eq.et}) with $\ee = \eet = 0.1$.
As shown in the left panel, the reconstructed jet structure for each light curve has the same shape as the counterpart in Figure~\ref{fig.result_star} but has a different total energy.
In fact, the relative difference of the jet shape $f(\theta)$ in Equation~(\ref{eq.shape}) is less than $10^{-6}$.\footnote{The relative difference between the jet structures comes from the numerical errors and the relativistic approximation ($\Gamma \gg 1$) used for deriving the degeneracy relation (See Appendix~\ref{sec.degen}). The situation is the same for the relative difference in the light curves.}
The total energy $\Etot$ scales with $n_0$, since $\Etot/n_0$ for each jet is the same as in Section~\ref{sec.results_n0-3} as presented in the rightmost column in Table~\ref{tab.result_bullet}.
As a result of the tuning, the total energy $\Etot$ is larger for spindle jets than for hollow-cone jets in Figure~\ref{fig.result_bullet}, which is contrast to the jets in Figure~\ref{fig.result_star}.

The top right panel of Figure~\ref{fig.result_bullet} shows the light curves generated by the jets in the left panel. We note that the light curves in Figures~\ref{fig.result_star} and \ref{fig.result_bullet} for each $a$ are almost indistinguishable.
In fact, the relative difference between the light curves in Figures~\ref{fig.result_star} and \ref{fig.result_bullet} is less than $10^{-6}$ in optical and X-ray.
This is a natural consequence from the degeneracy relation.

The above is also true in radio. However the relative difference is relatively large in early phase $T\sim 2$~d as displayed in Figure~\ref{fig.app_reladiff}, which shows that the relative difference is $\sim 0.1$ in the case of the spindle jet for $a=-10\times 10^{-3}$~d$^{-1}$ and the hollow-cone jet for $a=2\times 10^{-3}$~d$^{-1}$.
This is because the degeneracy relation holds only for the frequencies that lies in $\numd \le \nu' \le \nucd$ (See Appendix~\ref{sec.degen}) while $\nu' < \numd$ is realized for $\nu=5.5$~GHz in a region close to the jet axis when the jet Lorentz factor $\Gamma$ is still large. The frequency can drop below the synchrotron characteristic frequency due to the different dependencies of $\nu'$ and $\numd$ on $\Gamma$:
$\nu' \propto \Gamma \nu$ and $\numd \propto \gamma_\mathrm{m}'{}^2 B' \propto \Gamma^3$, where we assumed the observer direction is outside the beaming cone and $\gamma_\mathrm{m}'$ and $B'$ are the minimal Lorentz factor of the accelerated electrons and the magnitude of the magnetic field in the proper frame, respectively.
We can show that $\emisd$ increases with $n_0$ for a frequency below the characteristic frequency, $\nu' < \numd$, as given by Equation~(\ref{eq.emisdconst}). The early emission from the inner region is not significant because of the relativistic debeaming, but it leads to a quantitative difference as shown in Figure~\ref{fig.app_reladiff}. The radio light curves obtained here is brighter than those in the previous subsection for $a \le -4 \times 10^{-3}$ because $n_0$ is larger as shown in Figure~\ref{fig.eBn0}. On the other hand, the afterglow becomes dimmer in radio because of smaller $n_0$ for $a \ge -2 \times 10^{-3}$ than that in Section~\ref{sec.results_n0-3}. The region with $\nu' < \numd$ shrinks as the shock is decelerated and, hence, the difference between the light curves decreases as time passes.

As demonstrated above, the jet structure is determined from a given light curve except for the energy normalization. It is necessary to fix $n_0$, $\eB$, and $\ee$ for determining the energy scale (See Section~\ref{sec.dof} for the uncertainty of $\view$). 
Early observations with multiple frequencies across the synchrotron break frequency may resolve the degeneracy of the energy scaling.

\section{Discussions}\label{sec.discussion}
\subsection{Why do the jets with totally different structures generate similar light curve peaks?}\label{sec.discussion_similar}
We here discuss the result in Section~\ref{sec.results_n0-3}: Why do the jets with totally different structures generate afterglows with nearly the same peak time and flux, as presented in Figure~\ref{fig.result_star}?

\begin{figure}
	\includegraphics[width = \columnwidth]{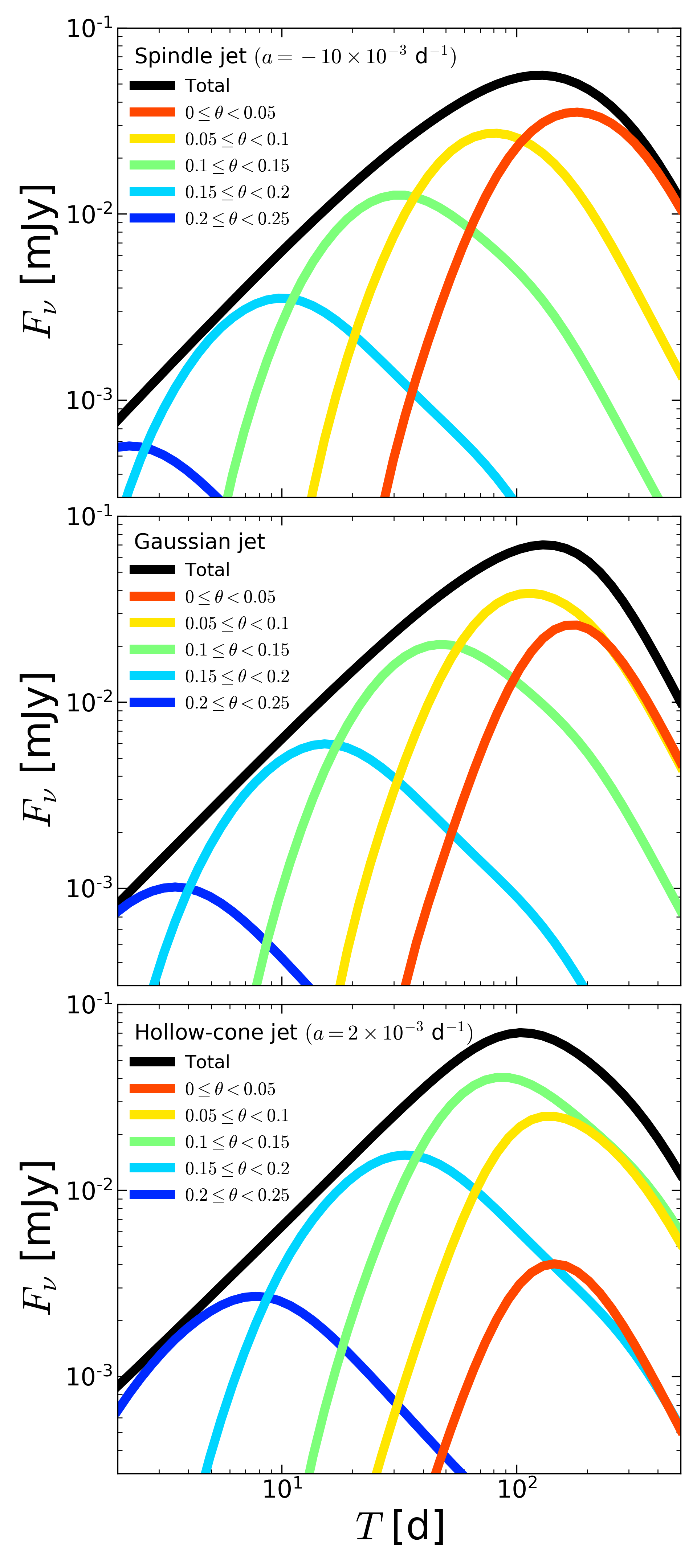}
	\caption{Decomposition of the radio light curve in Figure~\ref{fig.result_star} for the spindle jet obtained for $a=-10\times 10^{-3}$~d$^{-1}$, Gaussian jet, and hollow-cone jet obtained for $a=2\times 10^{-3}$~d$^{-1}$ from top to bottom, respectively. Each coloured line shows the light curve that is contributed from an annulus area on the jet surface, while the black one shows the total light curve, which is the same as in Figure~\ref{fig.result_star}.}
	\label{fig.result_decompLC}
\end{figure}

To answer the question, we decompose the radio light curve into the contributions from annulus regions on the jet surface, $\theta_n \le \theta \le \theta_{n+1}$ with $\theta_n = 0.05n$ ($n=0,1,2,\cdots$). Figure~\ref{fig.result_decompLC} shows the decomposition of the light curves generated by three representative jet structures in Figure~\ref{fig.result_star}: the spindle jet for the curvature parameter $a=-10\times 10^{-3}$~d$^{-1}$, the Gaussian jet, and the hollow-cone jet for $a=2\times 10^{-3}$~d$^{-1}$ from top to bottom, respectively.
In the case of the spindle jet, the more inner region gradually dominates the the light curve as time passes and the most inner region close to the jet axis contributes to the peak flux. This is natural, since the energy of the spindle jet is higher for more inner region. 
On the other hand, in the case of the hollow-cone jet, the peak of the light curve is dominated by the emission from the middle region of the jet with $0.1 \le \theta \le 0.15$ while the emission from the more inner region is sub-dominant. As shown in Figure~\ref{fig.result_star}, $E(\theta)$ peaks around $\theta \sim 0.1$ for the hollow-cone jet and the energy at $\theta = 0.1$ is larger than that for the spindle jet. Hence, recalling that the ambient density $n_0$ is the same, the middle region of the hollow-cone jet becomes visible later than that of the spindle jet.
The peak flux is adjusted by tuning $\eB$. The hollow-cone jet needs smaller $\eB$ than the spindle jet to suppress the emission from the middle region.
The emission coming from the central region is sub-dominant in the hollow-cone jet simply because the jet energy is lower than in the middle region. In the case of the Gaussian jet (middle panel), which lies between the spindle jet and the hollow-cone jet, the emission from the jet region with $0.05 \le \theta < 0.1$ produces the peak.

To sum up, the afterglow peaks can be similar even for totally different jet structures, because of the two reasons:
\begin{enumerate}
    \item The light curve peak is generated by the emission from the characteristic position of the jet, which is the energy peak on the middle region in the case of hollow-cone jets while it is the jet axis in the case of spindle jets.
    \item The emission from the characteristic position is adjustable by tuning the energy scale of the jet (or the number density $n_0$) and the energy conversion fraction from the shocked matter to the magnetic field $\eB$. The peak time $\Tp$ is adjusted by tuning the energy scale: The peak time is advanced/delayed for smaller/larger energy when $n_0$ is fixed. It is also possible to shift the peak time to an earlier/later time by giving smaller/larger $n_0$ \citep{NP20}. The peak flux is adjusted by $\eB$. Smaller $\eB$ reduces the synchrotron flux while larger $\eB$ enhances the emission.
\end{enumerate}

\subsection{Why does the curvature of the light curve slope determine the shape of the jet structure?}
As shown in Section~\ref{sec.results}, different jet structures are reconstructed from light curves with different values of the curvature parameter $a$, which controls the changing rate of the light curve slope.
As shown in \citetalias{TI}, we confirm that the jet structure is mainly determined by a given light curve while the other parameters only quantitatively change the structure. 
\citetalias{TI} also showed that the reconstructed structure does not qualitatively depend on a level of approximation by studying the dependence on $\fb$, where the calculated flux becomes more accurate for a larger $\fb$ since the inner cutoff angle $\Theta(T)$ becomes smaller for a given $T$ [See Equation~(\ref{eq.Theta})]. Hence, the qualitative difference among the jet structures originates in the difference of the curvature parameter. On the other hand, the resultant light curves are still so similar to each other that all of them are consistent with the observations within errors as shown in Figures~\ref{fig.result_star} and \ref{fig.result_bullet}.

The reason is explained as follows. The slope is determined by how rapidly the inner region of the jet becomes visible. The light curve goes upward with time (i.e., the light curve is convex downward) if the inner region becomes visible earlier. On the other hand, the light curve becomes convex upward if the inner region is seen later. For example, a spindle jet is reconstructed from a convex upward light curve with $a<0$. This is because the jet inner region is seen late because of the relativistic beaming with the higher jet energy contained around the jet axis. On the other hand, a convex-downward light curve with $a>0$ reconstructs a hollow-cone jet, because the jet inner region becomes visible earlier.

\section{Summary \& Conclusions}\label{sec.conclusion}
We investigated possible jet structures consistent with the afterglow of GRB~170817A by applying the method of the inverse reconstruction in our previous paper \citepalias{TI}. We studied how the jet structure depends on the light curve by gradually changing the curvature parameter of the light curve $a$ within the observational errors, which gives convex downward light curves for $a>0$ and convex upward ones for $a<0$. We found that hollow-cone jets are reconstructed from light curves with larger $a$ while spindle jets are reconstructed from light curves with smaller $a$. The structure gradually changes by passing through a Gaussian-like structure as $a$ changes.

In spite of the totally different jet structures, they generate similar light curves, which are consistent with the observed afterglow of GRB~170817A within errors. The light curve peak is produced by the emission from the jet surface where the jet energy peaks. The peak time and the peak flux can be adjusted for each jet structure by tuning the number density of the ambient medium $n_0$, the energy conversion fraction from the shocked matter to the magnetic field $\eB$, and the energy conversion fraction from the shocked matter to the accelerated electrons $\ee$.

These jet structures are distinguished by precisely observing the rising afterglow with the aid of the inversion method. The precision less than 18 per cent is required in the early phase $T\le 98$~d to distinguish the spindle jet from the Gaussian jet while less than 6 per cent is necessary in $T\le 37$~d to distinguish the hollow-cone jet from the Gaussian jet as indicated by Figure~\ref{fig.result_star}. In the case of GRB~170817A, the observational one-sigma errors are comparable or larger in the rising phase \citep{Panchromatic}. Hence, more precise observations are necessary for determining jet structures from the inverse reconstruction. Such observations would be possible for future events in denser environment, for which afterglow light curves become brighter \citep{GNP19,Duque20,OConnor20}. In the same reason, future events with a smaller viewing angle are also preferable, whereas the viewing angle has to be large enough so that the jet structure is reflected into observed light curves.

We found that the total energy of the jet can be tuned without changing the jet shape by using a different appropriate combination of the afterglow parameters ($n_0$, $\eB$, and $\ee$). We discovered that the degenerate combination is given by Equation~(\ref{eq.degen1}), which is applicable to any jet structure as demonstrated in this paper. We showed that the total energy is proportional to $n_0$. The degeneracy of the parameters $n_0$, $\eB$, and $\ee$ are broken if multi-frequency observations across the synchrotron break and cooling break are available and the viewing angle is fixed by other observations as mentioned below.
Here, future events in denser environment are preferred again, since the spectral break shifts to the observed frequency. Indeed, the synchrotron frequency and cooling frequency respectively satisfy $\nu_\mathrm{m}\propto E^{1/2}\eB^{1/2}\ee^2 T^{-3/2}$ and $\nu_\mathrm{c} \propto n_0^{-1}E^{-1/2}\eB^{-3/2}T^{-1/2}$ \citep{Sari98}. Increasing $n_0$ shifts $\nu_\mathrm{m}$ to a larger frequency and $\nu_\mathrm{c}$ to a smaller frequency, since the shock decelerates earlier and more inner region with larger $E$ is seen for larger $n_0$ when $\eB$, $\ee$, and $T$ are fixed.\footnote{The situation is the same even for hollow-cone jets as long as we consider the rising phase of the afterglow, since the light curve before the peak is generated by the emission from the region outside the energy peak as discussed in Section~\ref{sec.discussion_similar}.}

The viewing angle is another cause of the degeneracy, where a larger viewing angle leads to a wider jet shape \citep{NP20,Ryan20,TI}. As pointed out in \citet{NP20}, the degeneracy between the viewing angle and the jet-core angle is not broken only from the afterglow light curve but is resolved by an additional information such as the Lorentz factor inferred from the superluminal motion of the centroid.

As demonstrated in this paper, there is a large variety of possible jet structures for GRB~170817A consistent with the afterglow observations within errors. Since the inner jet structure is solely determined by the curvature of the light curve, qualitatively the same diversity should emerge even for the power-law edge as demonstrated in \citetalias{TI}.
There are several possible ways to form a hollow-cone jet in GRBs: the Blandford-Znajek mechanism \citep{BZ,McKinney06,TMN08}, the interaction between the jet and the ambient medium during the propagation through the ejecta or at the jet breakout \citep{Z03,MI13}, and jet precession \citep{McKinney13,Kawaguchi15,Huang19}, for example. It would be worth noting that hollow-cone jets are found in magnetohydrodynamical simulations \citep[e.g.,][]{Kathirgamaraju19,Nathanail20}, whereas
\citet{Gottlieb20} recently found jet structures consistent with a classical power-law jet in their hydrodyanamical simulations, which results from fluid instabilities at the jet-cocoon interface.
It is also worth noting that \citet{Salafia20} invoked a two-component jet similar to the spindle jet found in this paper that explains the luminosity function of the short GRBs, although they used a rather simple analytic model \citep[c.f.][]{Hamidani19,Hamidani20}.
Anyway, the jet structure of short GRBs is still under debate in theoretical studies.
High cadence and accuracy are necessary for the observations in the rising phase of the afterglow to determine the GRB jet structure.

\section*{Acknowledgements}
We thank Bing Zhang, Hamid Hamidani, Koutarou Kyutoku, Tomoki Wada, and Wataru Ishizaki for useful discussion. We also thank Ehud Nakar and Tsvi Piran for useful comments that have improved the manuscript.
We acknowledge the YITP workshops YITP-W-19-04, YITP-W-18-12, and YITP-W-18-11.
This work is supported by JSPS Grants-in-Aid for Scientific Research
17H06362 (KT, KI)
and 20H01904, 20H01901, 20H00158, 18H01213, 18H01215, 17H06357, 17H06131 (KI).

\section*{Data Availability}
The data underlying this article will be shared on reasonable request to the corresponding author.








\appendix
\section{Review of the inversion formula} \label{sec.review}
Based on \citetalias{TI}, we give a brief review of 
our inversion formula, Equation~(\ref{eq.inversion}), which is obtained by differentiating Equation~(\ref{eq.Fapp}) with respect to $T$ and using Equation~(\ref{eq.dThetadT}). See Section~\ref{sec.dof} for the basic picture of the afterglow model.

The observed afterglow is contributed only from a limited region of the jet surface around the line of sight of the off-axis observer due to relativistic beaming effects. Hence, the observed flux $F_\nu(T)$, where $\nu$ and $T$ are respectively the observed frequency and the observer time, is approximately given by integrating the synchrotron emission in a limited polar angle region between $\Theta(T)$ and $\thj$:
\begin{equation}
\label{eq.Fapp}
F_{\nu}(T) \sim \int_{\Theta(T)}^{\thj} \diff \theta K(T,\theta,E(\theta)),
\end{equation}
where $\thj$ is a given jet truncation angle. $\Theta(T)$ is the inner-side edge of the emission site that mainly contributes to the observed afterglow for each time, which is defined by
\begin{equation}
\label{eq.Theta}
\Theta = \view - \frac{\fb}{\Gamma} \sim \view - 4\fb \left[\frac{\pi n_0 \mpr c^5}{17(1 + 8\fb^2)^3}\right]^{1/8} E^{-1/8}(\Theta) T^{3/8},
\end{equation}
where $\mpr$ and $c$ are the proton mass and the speed of light, respectively. $\fb$ 
is the size of the emission region measured by $1/\Gamma$,
and we adopt $\fb=7$ throughout the paper as in \citetalias{TI}.
The inversion process is finished when $\Theta(T)$ becomes zero, which takes place at $\Tf$ given by
\begin{align}
    \label{eq.Tf}
    \Tf = 60.2\ \mathrm{day} \left(\frac{\Eax}{10^{52}\ \mathrm{erg}}\right)^{1/3}\left(\frac{n_0}{10^{-3}\ \mathrm{cm}^{-3}}\right)^{-1/3}\left(\frac{\view}{0.4} \right)^{8/3},
\end{align}
where $\Eax = E(0)$.
The time derivative of $\Theta$ is obtained by differentiating Equation~(\ref{eq.Theta}) with respect to $T$ as follows:
\begin{align}
\label{eq.dThetadT}
\frac{\diff \Theta}{\diff T} = -\frac{3(\view - \Theta)}{8T}\left(1 - \frac{\view - \Theta}{8}\frac{\diff \ln E}{\diff \Theta}\right)^{-1}.
\end{align}
In Equation~(\ref{eq.Fapp}), $K(T,\Theta,E(\Theta))$ is given by
\begin{align}
\label{eq.K}
K(T, \theta, E(\theta)) 
= \frac{1}{4\pi D^2} \int _0^{2\pi} \diff \phi \left. \frac{\sin \theta \Rs^3\emisd}{12\Gamma^4(1 - \beta_\sh \mu)(1 - \beta \mu)^2} \right| _{t = \ts},
\end{align}
which integrates the synchrotron emissivity $\emisd$ on a thin annulus. In the above equation, $\phi$ is the azimuthal angle around the jet axis and 
\begin{equation}
\label{eq.mu}
\mu = \sin \theta \sin \view \cos \phi + \cos \theta \cos \view
\end{equation}
is the cosine of the angle spanned by the radial vector and the line of sight for the observer located at $\phi=0$.
$\Gamma$ and $\beta$ are the Lorentz factor and the normalized speed of shocked fluid while those with the subscript `sh' are of the shock wave, which satisfy
\begin{align}
\label{eq.Gammabetash}
\Gamma_\sh^2 \beta_\sh^2 &= C_\BM^2 t^{-3},\\
\label{eq.Gammabeta}
\Gamma^2 \beta^2 &= \frac{1}{2}C_\BM^2 t^{-3},\\
\label{eq.C_BM}
C_\BM &= \sqrt{\frac{17E}{8\pi n_0\mpr c^5}},
\end{align}
with $t$ being the laboratory time elapsed from the explosion that launches the jet at the origin.
Equations~(\ref{eq.Gammabetash}) and (\ref{eq.Gammabeta}) reduce to a self-similar solution of \citet{BM} in the ultra-relativistic limit ($\beta = \beta_\sh = 1$). These factors, $\beta$ and $\beta_\sh$, are introduced to ensure $\Gamma _\sh>1$ and $\Gamma>1$ for any $t$, while they do not affect the observed flux, which is contributed from the relativistic region during the inversion process. $\Rs$ in Equation~(\ref{eq.K}) is the shock radius given by integrating $c\beta_\sh$ with respect to $t$:
\begin{equation}
\label{eq.Rs}
\Rs(t) = {}_2F_1\left(\frac{1}{3},\frac{1}{2},\frac{4}{3};-\frac{t^3}{C_\BM^2}\right) ct,
\end{equation}
where ${}_2F_1(\cdots)$ is the Gauss's hypergeometric function. The synchrotron emissivity $\emisd$ in Equation~(\ref{eq.K}) is given by
\begin{equation}
\label{eq.emisd}
\emisd = \epsilon'_{\nu',\mathrm{p}} \left( \frac{\nu '}{\numd}\right)^{-(p-1)/2},
\end{equation}
where we assume the observed frequency in the fluid rest frame 
\begin{align}
\label{eq.nud}
\nu'=\Gamma(1-\beta \mu)\nu	
\end{align}
lies between the characteristic frequency $\numd$ and the cooling frequency $\nucd$ as $\numd < \nu' < \nucd$, which is appropriate in the case of GRB~170817A. The characteristic frequency and the corresponding emissivity are given by \citep{Granot99,Eerten10}
\begin{align}
\label{eq.numd}
\numd &=  \frac{3}{16} \left[ \ee \frac{p-2}{p-1}\frac{\mpr}{\me}(\Gamma - 1) \right]^2 \frac{\qe B'}{\me c}, \\
\label{eq.emisdpeak}
\epsilon '_{\nu',\mathrm{p}} &= 0.88 \cdot \frac{256}{27}\frac{p-1}{3p-1}\frac{\qe^3}{\me c^2} n'B',
\end{align}
where $\qe$ is the elementary charge and $\me$ is the electron mass. $n'$ and $B'$ denote the number density and magnetic field measured at the fluid rest frame, which are given by
\begin{align}
\label{eq.nd}
n' &= 4\Gamma n_0, \\
\label{eq.Bd}
B' &= \sqrt{32\pi \eB n_0 \Gamma(\Gamma -1)\mpr c^2}.
\end{align}
$\ts$ in Equation~(\ref{eq.K}) is the laboratory time that emits the photons arriving at the observer at $T$, which is given as the solution of the following equation:
\begin{equation}
\label{eq.t}
\ts = T + \frac{\mu \Rs(\ts)}{c},
\end{equation}
where we choose $T=0$ as the arrival time of a photon emitted at the origin at $t=0$.
We employ an approximated analytic solution of Equation~(\ref{eq.t}) for $\ts$, which is obtained in the relativistic limit (See appendix A in \citetalias{TI}).

\section{Constraints on the jet edge structure} \label{app.jetedge}
We constrain the jet edge structure as follows.
Suppose that a light curve starts at an initial time $T_0$ and the parameters $\{n_0, \eB, \ee, \view, p, D\}$ and the jet truncation angle $\thj$ are given.
Then, the jet edge structure $E(\theta)$ $(\Theta(T_0)) < \theta \le \thj)$ should satisfy Equation~(\ref{eq.Fapp}) at the initial time $T_0$: 
\begin{align}
\label{eq.constraint1}
F_\nu(T_0) = \int _{\Theta(T_0)}^{\thj} \diff \theta K(T_0, \theta, E(\theta)).
\end{align}
In addition, if the light curve is smooth at $T_0$, the jet edge structure also satisfies the first derivative of Equation~(\ref{eq.constraint1}) at $T_0$:
\begin{align}
\nonumber
\frac{\diff F_\nu}{\diff T}(T_0) &= \frac{\diff \Theta}{\diff T}(T_0) K(T_0, \Theta(T_0), E(\Theta_0))\\
\label{eq.constraint2}
&\quad + \int _{\Theta(T_0)}^{\thj} \diff \theta \frac{\diff K}{\diff T}(T_0, \theta, E(\theta)) .
\end{align}
In the same manner, higher derivatives of the light curve at the initial time give constraints on $E(\theta)$ $(\Theta(T_0)) < \theta \le \thj)$ in principle, whereas the higher derivatives are difficult to obtain from observations.
In this paper, we use only Equations~(\ref{eq.constraint1}) and (\ref{eq.constraint2}). 
Since we assume a Gaussian edge given by Equation~(\ref{eq.GaussianJet}), these equations determine the normalization $E_0$ and the standard deviation $\thc$ of the jet edge.
See \citetalias{TI} for the prescription for the power-law edge.

\section{Derivation of the Degeneracy Relation} \label{sec.degen}
Let us derive the degeneracy relation described in Section~\ref{sec.degeneracy_main}.
That is, we show that for the combinations of $(n_0, \eB, \ee)$ in Equation~(\ref{eq.degen1}),
the jet structures reconstructed from a given light curve have the same shape $f(\theta)$ but the energy $\Eax$ scales with $n_0$ [Equation~(\ref{eq.degen2})], where we decompose a jet structure to the shape and normalization as given by Equation~(\ref{eq.shape}).

The inversion formula, Equation~(\ref{eq.inversion}), is equivalent to Equation~(\ref{eq.Fapp}). In the relativistic limit $(\Gamma \gg 1)$, which is a good approximation unless the shock is entirely decelerated to non-relativistic speeds, Equation~(\ref{eq.Fapp}) reduces to
\begin{align}
\label{eq.Fnupropto}
F_\nu(T) \propto [\Eax f(\Theta(T))]^{(p+3)/4}n_0^{1/2}\eB^{(p+1)/4}\ee^{p-1}T^{-3(p-1)/4}.
\end{align}
Here, we used  
$\Gamma \propto E^{1/2}n_0^{-1/2} t^{-3/2}$ [Equation~(\ref{eq.Gammabeta})] and
$t \sim T/(1 - \beta \mu)\sim \Gamma^2 T$ [Equation~(\ref{eq.t})], which lead to $\Gamma \propto E^{1/8}n_0^{-1/8}T^{-3/8}$. We also employed 
$\Rs \sim c \beta t \propto \Gamma^2T $ [Equation~(\ref{eq.Rs})],
$\emisd \propto n_0^{(p+5)/4} \eB^{(p+1)/4} \ee ^{p-1} \Gamma ^{p+1} (1-\beta \mu)^{-(p-1)/2} \propto n_0^{(p+5)/4} \eB^{(p+1)/4} \ee ^{p-1} \Gamma ^{2p}$ [Equation~(\ref{eq.emisd})],
and $\Delta \Omega \propto \Gamma^{-2}$, where $\Delta \Omega$ is the solid angle of the luminous region that dominates the integral.
$\Theta(T)$ in Equation~(\ref{eq.Fnupropto}) obeys Equation~(\ref{eq.Theta}). Equation~(\ref{eq.Theta}) can be also written by
\begin{align}
\label{eq.Thetapropto}
\Theta(T) = \left\{
    \begin{array}{ll}
        \view - A \left[\frac{\Eax f(\Theta(T))}{n_0}\right]^{-1/8}T^{3/8} & (T \le \Tf )\\
        0 & (T > \Tf)
    \end{array}\right.,
\end{align}
where $A$ is a constant.
Eliminating $\Eax f(\Theta(T))$ from Equation~(\ref{eq.Fnupropto}) with Equation~(\ref{eq.Thetapropto}), we obtain
\begin{align}
    \label{eq.Fnupropto_reduced}
    F_\nu(T) \propto [\view - \Theta(T)]^{-2(p+3)}n_0^{(p+5)/4}\eB^{(p+1)/4}\ee^{p-1}T^3.
\end{align}
Since a light curve $F_\nu(T)$ and the viewing angle $\view$ are supposed to be fixed, the following things should hold: First, $n_0^{(p+5)/4}\eB^{(p+1)/4}\ee^{p-1}$ is a constant; Second, $\Theta(T)$ is a fixed function of $T$. The former is equivalent to the degeneracy relation, Equation~(\ref{eq.degen1}). The latter and Equation~(\ref{eq.Thetapropto}) lead to $\Eax \propto n_0$, Equation~(\ref{eq.degen2}), and that $f(\theta)$ is a fixed function. Conversely, if $f(\theta)$ is fixed and Equations~(\ref{eq.degen1}) and (\ref{eq.degen2}) are satisfied, $\Theta(T)$ is an invariant and hence $F_\nu(T)$ is also an invariant.

Note that we assumed the synchrotron emissivity is expressed by Equation~(\ref{eq.emisd}) in the above. Hence, the degeneracy relation holds only for the frequencies in $\numd \le \nu' \le \nucd$.

\section{Synchrotron emissivity for $\nu' < \numd$}\label{app.ii}
We here argue the synchrotron emissivity $\emisd$ for $\nu' < \numd$ in the slow cooling case, which is used for the discussion in Section~\ref{sec.results_another}. We assume Equations~(\ref{eq.degen1}) and (\ref{eq.degen2}) and use the relativistic limit ($\Gamma \gg 1$). For a fixed angular coordinate $(\theta, \phi)$ on the jet surface, Equations~(\ref{eq.nud}) and (\ref{eq.numd}) lead to
\begin{align}
	\label{eq.app1}
	\frac{\nu'}{\numd} \propto \Gamma^{-2} n_0^{-1/2}\eB^{-1/2}\ee^{-2}.
\end{align}
The Lorentz factor given by Equation~(\ref{eq.Gammabeta}) obeys the following proportionality for a fixed $(\theta, \phi)$:
\begin{align}
	\label{eq.app2}
	\Gamma \propto E^{1/2} n_0^{-1/2} T^{-3/2},
\end{align}
where we used $t\sim T/(1 - \mu \beta) \propto T$ for fixed $\mu$.
Substituting Equations~(\ref{eq.degen1}), (\ref{eq.degen2}), and (\ref{eq.app1}) to Equation~(\ref{eq.app2}), we obtain
\begin{align}
	\label{eq.break}
	\frac{\nu'}{\numd} &\propto n_0^{2/(p+1)} \ee^{-4/(p+1)}T^3.
\end{align}
In the same manner, we obtain the following equation from Equations~(\ref{eq.degen1}), (\ref{eq.degen2}), (\ref{eq.emisdpeak}), and (\ref{eq.app2}):
\begin{align}
	\label{eq.peak}
	\epsilon '_{\nu',\mathrm{p}} &\propto n_0^{(p-1)/(p+1)} \ee^{-2(p-1)/(p+1)} T^{-3}.
\end{align}
Taking into account that the power-law index for $\nu' < \numd$ is $1/3$ in the case of the slow cooling \citep{Sari98}, we obtain the following equation from Equations~(\ref{eq.break}) and (\ref{eq.peak}):
\begin{align}
	\label{eq.emisdconst}
	\emisd \propto 
	n_0^{(3p-1)/[3(p+1)]} \ee^{2(3p-1)/[3(p+1)]}T^{-2}\ (\nu' < \numd) .
\end{align}
Thus, if $p>1/3$, the synchrotron emissivity decreases for smaller $n_0$ and $\ee$ for the frequencies below the characteristic frequency, $\nu' < \numd$.


\bsp	
\label{lastpage}
\end{document}